\documentclass[aps,twocolumn,preprintnumbers,amsmath,nofootinbib,amssymb]{revtex4}

\usepackage{graphics}
\usepackage{epsfig,epsf,psfrag}
\usepackage{times}
\usepackage{float}
\usepackage{color}
\usepackage{amsfonts,amssymb,stmaryrd,latexsym,amsmath}
\usepackage{multirow}
\usepackage{array}
\usepackage{hhline}
\usepackage{booktabs}
\usepackage{enumerate}
\usepackage{slashed}
\usepackage{subfigure}
\usepackage[colorlinks, citecolor=blue,anchorcolor=red,menucolor=red, linkcolor=red,filecolor=red,runcolor=red,urlcolor=blue,frenchlinks=red]{hyperref}
\usepackage{epstopdf}

\begin{document}

\bibliographystyle{unsrt}

\title{Possibility of $T_{c\bar{s}}(2900)$ as the resonance-like structure induced by threshold effects}

\author{Ying-Hui~Ge$^{1}$}
\author{Xiao-Hai~Liu$^{1}$}~\email{xiaohai.liu@tju.edu.cn}
\author{Hong-Wei Ke$^{1}$}~\email{khw020056@tju.edu.cn}

\affiliation{ $^1$Center for Joint Quantum Studies and Department of Physics, School of Science, Tianjin University, Tianjin 300350, China	}

\date{\today}

\begin{abstract}

We investigate the process $B\to \bar{D}D_s \pi$ via several rescattering processes. It is shown that the triangle singularity (TS) peak around the $D^*K^*$ threshold generated from the $\chi_{c1}K^* D^*$ loop is relatively narrow, which may simulate the resonance-like structure $T_{c\bar{s}}(2900)$ recently observed by LHCb in the $D_s\pi$ spectrum. However, the TS peak around the $D_s^*\rho$ threshold generated from the $D^{**} D_s^* \rho$ loop is smoothed by the broad width of $\rho$, which itself can hardly describe the $T_{c\bar{s}}(2900)$ structure. A TS signal around the $DK$ threshold generated from the $\chi_{c0}K D $ loop is also predicted.


\end{abstract}

\maketitle

\section{Introduction}
Very recently, the LHCb collaboration reported the observation of two new tetraquark candidates in the $B^0\to \bar{D}^0 D_s^+ \pi^-$ and $B^+\to D^- D_s^+\pi^+$ decays~\cite{LHCb:Tcs2900}. Their masses and widths are
\begin{eqnarray}\label{Xstates}
T_{c\bar{s}}(2900)^0: && 	M=2892 \pm ~21 \mbox{MeV},
	\Gamma= 119\pm 29~  \mbox{MeV},  \nonumber\\
 \ T_{c\bar{s}}(2900)^{++}: &&	M=2921 \pm ~23 \mbox{MeV}, 
	\Gamma= 137\pm 35~  \mbox{MeV} \nonumber.
\end{eqnarray}
 Supposing they belong to the same isospin triplet, the experiment also gives the shared values
\begin{equation}
M=2908 \pm ~23 \mbox{MeV}, 
	\Gamma= 136\pm 25~  \mbox{MeV} \nonumber.
\end{equation}
The preferred spin-parity quantum numbers of $T_{c\bar{s}}(2900)^0$ and $T_{c\bar{s}}(2900)^{++}$ are $0^+$.
These two new members of the $XYZ$ particle family have striking features. Since they are observed in the $D_s^+\pi^-$ and $D_s^+\pi^+$ invariant mass spectrum, their valence quark contents are supposed to be $c\bar{s}d\bar{u}$ and $c\bar{s}u\bar{d}$, respectively. Although tens of exotic hadron candidates have been discovered since 2003, the fully open-flavor tetraquark states are still very rare~\cite{PDG:2022}. 
In 2016 the D0 collaboration ever reported the observation of a state $X(5568)$ in the $B_s^0\pi^\pm$ spectrum~\cite{D0:2016mwd}, which was then thought to be a fully open-flavor tetraquark state with the quark contents $\bar{b}s\bar{d}u$ (or $\bar{b}s\bar{u}d$). But its existence was not confirmed in the LHCb experiment~\cite{Aaij:2016iev}. The existence of $X(5568)$ was also severely challenged on theoretical grounds in consideration of its low mass~\cite{Burns:2016gvy,Guo:2016nhb,Kang:2016zmv}. There are some discussions on the possible reason of its appearance in the D0 and absence in LHCb and CMS in Ref.~\cite{Yang:2016sws}. In 2020, the LHCb collaboration reported two fully open-flavor tetraquark candidates $X_0(2900)$ and $X_1(2900)$ in $B^+\to D^+ D^-K^+$ decays~\cite{LHCb:2020bls,LHCb:2020pxc}. Since they are observed in the $D^- K^+$ spectrum, their valence quark contents are supposed to be $\bar{c}\bar{s}du$.
Concerning the nature of $X(2900)$, there have been many interpretations, such as the compact tetraquark states~\cite{Wang:2020xyc,He:2020jna,Zhang:2020oze,Wang:2020prk}, the hadronic molecule states composed of $D^*\bar{K}^*$ or $D_1 \bar{K}$~\cite{Liu:2020nil,Chen:2020aos,Huang:2020ptc,Molina:2020hde,Xue:2020vtq,Lu:2020qmp,Agaev:2020nrc,Mutuk:2020igv,Xiao:2020ltm,He:2020btl}, threshold effects~\cite{Liu:2020orv,Burns:2020epm}, and so on. The newly observed $T_{c\bar{s}}(2900)$ has very similar mass and quark contents with those of $X(2900)$. It would be natural to think that they maybe have the similar origin, and the $T_{c\bar{s}}(2900)$ and $X(2900)$ could be in partnership with each other. There have been some earlier studies concerning such open-flavor states in Refs.~\cite{Chen:2017rhl,Agaev:2022duz,Lu:2020qmp,He:2020jna,Azizi:2018mte,Cheng:2020nho,Albuquerque:2020ugi,Guo:2021mja,An:2022vtg}

We named the hadronic molecule, tetraquark or pentaquark state interpretation of those exotic hadron candidates as the genuine resonance interpretation. Besides, it has been shown that in some situation the kinematic singularities of rescattering amplitudes, such as the two-body threshold singularity and the triangle singularity (TS), can also generate resonance-like peaks in pertinent invariant mass spectra, which implies the non-resonance interpretation for some $XYZ$ states is possible. Before claiming that one resonance-like peak corresponds to one genuine particle, it is also necessary to exclude or confirm these possibilities. There have been quite a few exotic phenomena that are suggested to be induced by the threshold effects. We refer to Ref.~\cite{Guo:2019twa} for a recent review about the threshold cusp and TS in hadronic reactions.

In Ref.~\cite{Liu:2020orv}, we investigate the rescatterings which may play a role in $B^+\to D^+ D^- K^+$ decays. It is shown that the $D^{*-}K^{*+}$ rescattering via the $\chi_{c1}K^{*+}D^{*-}$ loop and the  $\bar{D}_{1}^{0}K^{0}$ rescattering via the $D_{sJ}^{+}\bar{D}_{1}^{0}K^{0}$ loop can mimic the $X_0(2900)$ and $X_1(2900)$ with consistent quantum numbers. A similar mechanism was also discussed in Ref.~\cite{Burns:2020epm}. Such phenomena are due to the analytical property of the scattering amplitudes with the TS located to the vicinity of the physical boundary. Taking into account the similarity between $X(2900)$ and $T_{c\bar{s}}(2900)$, we expect the similar mechanism may also work in explaining the observation of $T_{c\bar{s}}(2900)$.

\section{The Model}
\begin{figure}[htb]
    \centering
    \includegraphics[width=0.8\linewidth]{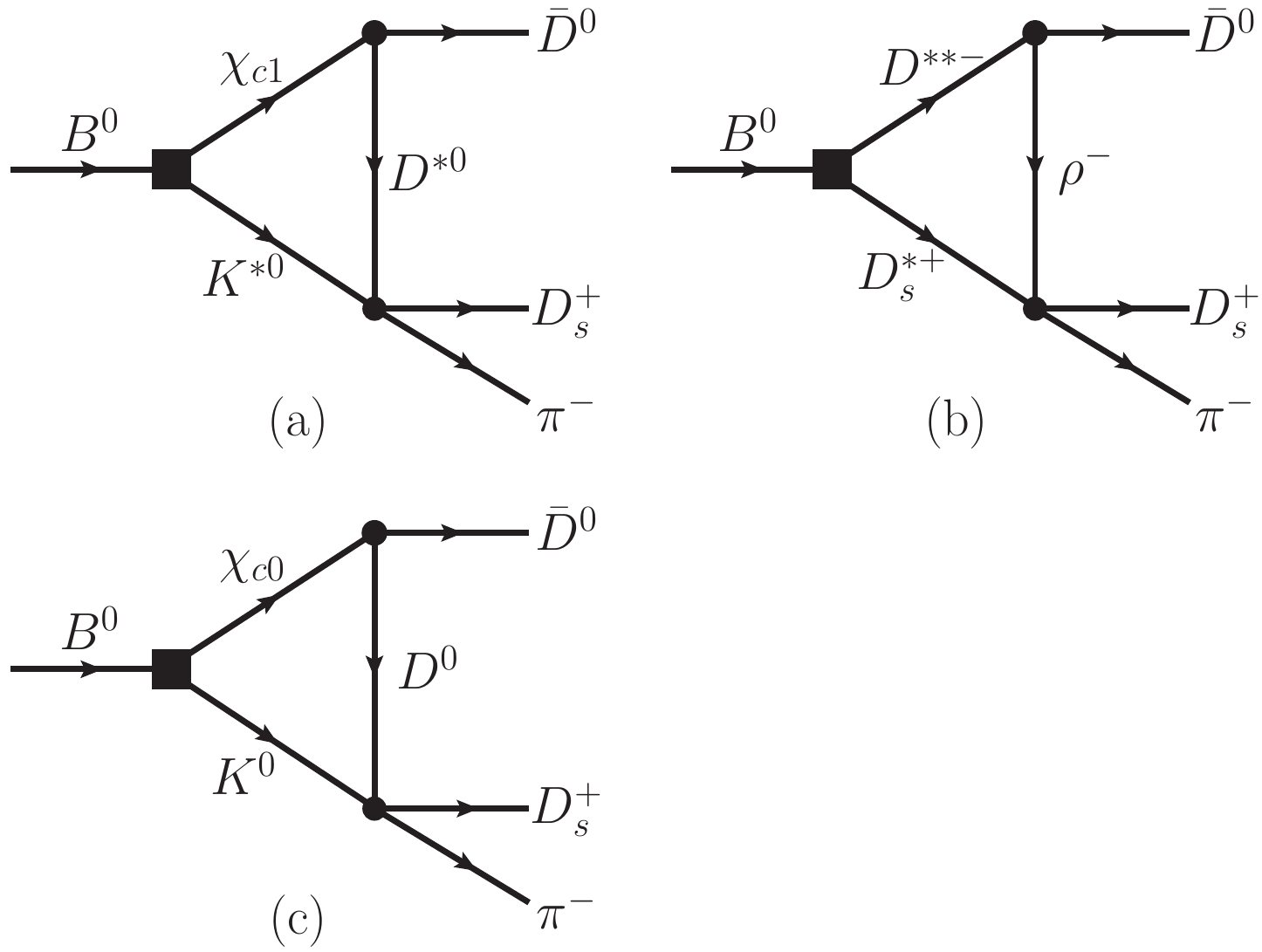}
    \caption{Rescattering diagrams which contribute to $B^0\to \bar{D}^0 D_s^+ \pi^-$. Kinematic conventions for the intermediate states are (a) $K^{*0}(q_1, m_1)$, $\chi_{c1}(q_2, m_2)$, $D^{*0} (q_3, m_3)$, (b) $D_s^{*+}(q_1, m_1)$, $D^{**}(q_2,m_2)$, $\rho^{-} (q_3, m_3)$, and (c) $K^{0}(q_1, m_1)$, $\chi_{c0}(q_2, m_2)$, $D^{0} (q_3, m_3)$, respectively.}
    \label{feynman-diagram}
\end{figure}

The $B^0$ meson decaying into a charmonium and a kaon, or a charmed-strange meson and an anti-charmed meson is the Cabibbo-favored process. Therefore it is expected that the rescattering processes illustrated in Fig.~\ref{feynman-diagram} may play a role in the decay $B^0\to \bar{D}^0 D_s^+ \pi^-$. The intermediate state $\chi_{c1}$/$\chi_{c0}$ in Fig.~\ref{feynman-diagram}(a)/(c) represents any charmonia with $J^{PC}=1^{++}$/$0^{++}$. The $D^{**-}$ state in Fig.~\ref{feynman-diagram}(b) represents any charmed meson with $J^{P}=1^{+}$. We only take into account the $D_s^+\pi^-$ in relative $S$-wave, which implies the quantum numbers of $D_s^+\pi^-$ system are $0^+$. In order to keep the conservation of angular momentum, we have above requirements on the quantum numbers of intermediate states. All of the three vertices $\chi_{c1}\to \bar{D}^0 D^{*0}$, $D^{**-}\to \bar{D}^0 \rho^-$ and $\chi_{c0}\to \bar{D}^0 D^{0}$ are $S$-wave couplings. There are a series of experimentally established and theoretically predicted charmonia and charmed meson with the required quantum numbers~\cite{PDG:2022}.

Another intriguing feature of the rescattering triangle diagrams illustrated in Fig.~\ref{feynman-diagram} is that the $ K^{*0}\chi_{c1}$, $D_s^{*+}D^{**-}$ or $K^0\chi_{c0}$ threshold could be close to $M_{B^0}$. As a result the TS of the rescattering amplitude is expected to locate near the physical boundary. The TS then may enhance the two-body threshold cusp or itself may generate a resonance-like peak in the $D_s^+\pi^-$ spectrum. The thresholds of $D^{*0}K^{*0}$ and $D_s^{*+}\rho^-$ are about 2902 MeV and 2887 MeV respectively, which are close to the mass of $T_{c\bar{s}}(2900)$. It is then expected that the nearby TSs corresponding to Figs.~\ref{feynman-diagram}(a) and (b) may mimic the $T_{c\bar{s}}(2900)$ structure.
As for Fig.~\ref{feynman-diagram}(c), the resonance-like structure induced by $D^0 K^0\to D_s^+\pi^-$ rescattering around $D^0 K^0$ threshold can also be expected.

\begin{figure}[htb]
	\centering
	\includegraphics[width=0.8\linewidth]{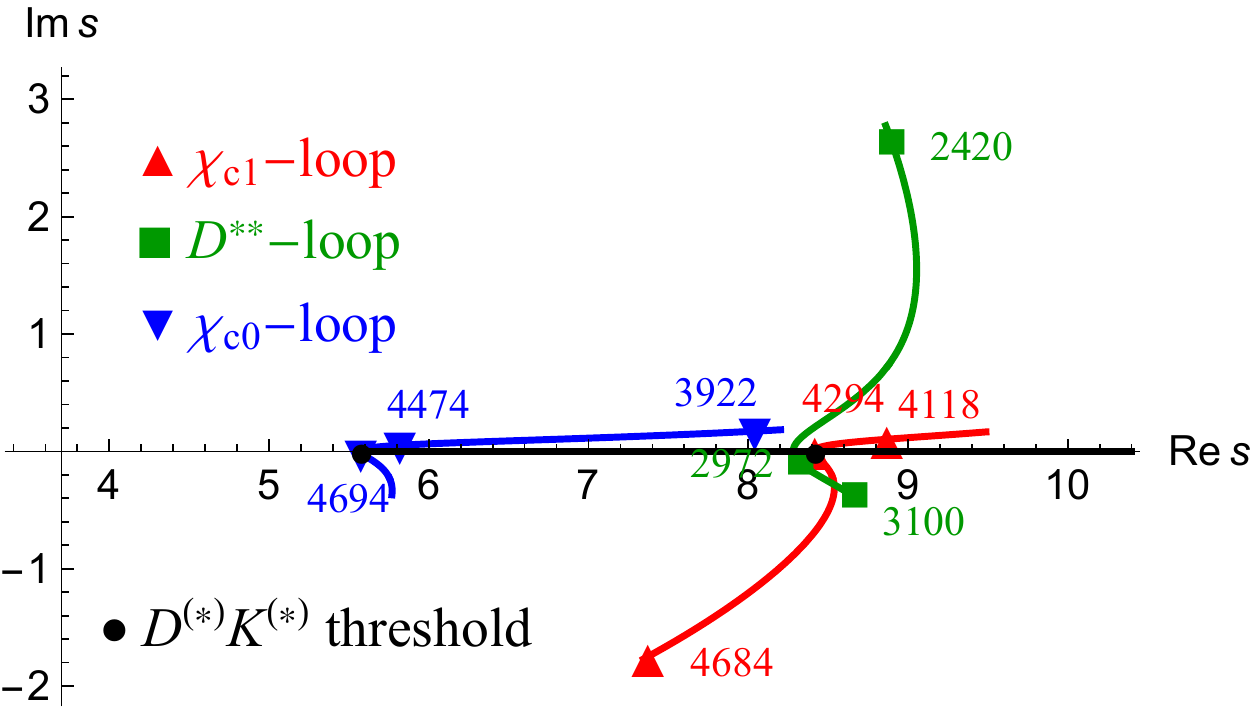}
	\caption{The TS location of the rescattering amplitude on the complex $s$-plane (in unit of GeV$^2$). The thick line on the real axis represents the unitarity cut starting from the $DK$ and $D^*K^*$($D_s^*\rho$) thresholds. The trajectories marked with up triangle, box, down triangle are obtained by varying $M_{\chi_{c1}}$, $M_{D^{**}}$ and $M_{\chi_{c0}}$ respectively. The widths of intermediate states are fixed to be $\Gamma_{\chi_{c1}}=50\ \mbox{MeV}$, $\Gamma_{D^{**}}=150\ \mbox{MeV}$, and $\Gamma_{\chi_{c0}}=50\ \mbox{MeV}$, respectively.  The numbers beside the markers represent the masses of corresponding intermediate state (in unit of MeV).}
	\label{TS-position}
\end{figure}

We define the $D_s^+\pi^-$ invariant mass square
$s\equiv M_{D_s^+\pi^-}^2$. The rescattering
amplitude corresponding to
Fig.~\ref{feynman-diagram}(a)/(b)/(c) has a normal threshold singularity
at the  $D^{*}K^{*}$/$D_s^{*}\rho$/$DK$ threshold, which is the start
point of the right-hand unitarity cut on the complex $s$-plane. The
cut leads to a two-sheet structure for the amplitude, and the
physical region is above the real axis on
the first Riemann sheet. This unitarity cut results in a threshold cusp at the
$D_s^+\pi^-$ distribution curve. With some special kinematic configurations, all of
the three intermediate states in Fig.~\ref{feynman-diagram}
can be on-shell simultaneously. In such a case, the amplitude has a leading Landau singularity, which is called the TS. The TS is found to be
located on the second Riemann sheet
\cite{Bronzan:1963mby,Aitchison:1964zz,Schmid:1967ojm}. If the TS of the amplitude in the complex $s$-plane is close to the physical
boundary, the physical amplitude may feel its influence.

The TS is a logarithmic singularity. To avoid the infinity of the
loop integral in the physical region, one can replace the Feynman's
$i\epsilon$ for the $m_2$ propagator by $im_2 \Gamma_2$ with
$\Gamma_2$ the total decay width, or equivalently replace the real
mass $m_2$ by the complex mass
$m_2-i\Gamma_2/2$~\cite{Aitchison:1964rwb}, which will remove the TS
from the physical boundary by a distance. The physical meaning of this
prescription for dealing with the infinity is obvious: as long as the
kinematic conditions for the TS being present on the physical
boundary are satisfied, it implies that the intermediate
state with the mass $m_2$ is unstable, and it is then necessary to take into account the
width effects.

For the triangle diagrams shown in Fig.~\ref{feynman-diagram}, the location of the TS in $s$ is given by \cite{Landau:1959fi,Coleman:1965xm,Bronzan:1963mby,Aitchison:1964zz,Liu:2015taa}
\begin{eqnarray}\label{eq:TSlocation}
s_{TS}&=&(m_1+m_3)^2+\frac{1}{2m_2^2} {\LARGE[}(m_2^2+m_3^2-M_{\bar{D}^0}^2) \nonumber \\  &\times&(M_{B^0}^2-m_1^2-m_2^2)-4m_2^2 m_1 m_3 \nonumber \\ &-& \lambda^{1/2}(M_{B^0}^2,  m_1^2,  m_2^2)\lambda^{1/2}(m_2^2,m_3^2,M_{\bar{D}^0}^2){\LARGE ]},
\end{eqnarray}
with $\lambda(x,y,z)= (x-y-z)^2- 4yz$. The TS location of rescattering amplitudes corresponding to Figs.~\ref{feynman-diagram}(a), (b) and (c) is displayed in Fig.~\ref{TS-position}. The three trajectories show the movement of TS location when the mass $m_2$ of the intermediate state varies, with other particle masses fixed.
In terms of Eq.~(\ref{eq:TSlocation}), the TS for the diagrams shown in Fig.~\ref{feynman-diagram} is on the physical boundary when $m_2^2$ falls in the range:
\begin{eqnarray}
\frac{m_1 M_{\bar{D}^0}^2+ m_3 M_{B^0}^2}{m_1+m_3} -m_1 m_3\leq m_2^2\leq(M_{B^0}-m_1)^2,
\end{eqnarray}
corresponding to the range
\begin{eqnarray}
(m_1+m_3)^2 &\leq& s_{TS} \leq(m_1+m_3)^2 \nonumber \\
&+&\frac{m_1[(m_2-m_3)^2-M_{\bar{D}^0}^2] }{m_2}.
\end{eqnarray}
Inputting the physical particle masses from Ref.~\cite{PDG:2022}, we obtain the range of $m_2$ that satisfies the requirements of TS being located on the physical boundary, which is displayed in Table~\ref{table:TSregion}. A series of intermediate states with proper quantum numbers can also (nearly) fulfill the mass requirement of the TS. We will introduce these states later. The explicit rescattering contributions from Figs.~\ref{feynman-diagram}(a), (b) and (c) are discussed in the following subsections separately.

For the sake of brevity, we focus our discussion on the $B^0\to \bar{D}^0 D_s^+ \pi^-$ process. The theoretical analysis and numerical results for $B^+\to D^- D_s^+\pi^+$ process are similar.

\begin{table}[tb]
	\renewcommand\arraystretch{1.2}
		\begin{ruledtabular}
 	\caption{TS kinematic region (all masses are real) corresponding to the rescattering diagrams in Fig.~\ref{feynman-diagram}, in unit of MeV.}\label{table:TSregion}
 		\begin{tabular}{c|c|c}
 			Diagram & $ m_2 $ & $ \sqrt{s_{TS}} $  \\
 			\hline
 			Fig.~\ref{feynman-diagram}(a)&$\chi_{c1}$: 4307$\sim$4384 & 2902$\sim$2978   \\
 			\hline
 			Fig.~\ref{feynman-diagram}(b)&$D^{**}$: 2897$\sim$3167&  2887$\sim$3136  \\
 			\hline
			Fig.~\ref{feynman-diagram}(c)&$\chi_{c0}$: 4670$\sim$4782&   2362$\sim$2471  \\
 		\end{tabular}
 	\end{ruledtabular}
 \end{table}

\subsection{$\chi_{c1}K^* D^*$ loop}
In Fig.~\ref{feynman-diagram}(a), if the $B^0$ mass is close to the $\chi_{c1}K^{*0}$ threshold, the $S$-wave decay is expected to be dominated. The general $S$-wave decay amplitude can be written as
\begin{eqnarray}\label{B0tochic1}
	\mathcal{A}(B^0\to \chi_{c1}K^{*0}) &=& w_a \epsilon^*_{\chi_{c1}}\cdot \epsilon^*_{{K}^{*0}},
\end{eqnarray}
where $w_a$ represents the weak coupling constant.

For the process $\chi_{c1}\to \bar{D}^0 D^{*0}$, the $S$-wave amplitude reads
\begin{eqnarray}\label{chic1toDDstar}
	\mathcal{A}(\chi_{c1}\to \bar{D}^0 D^{*0}) &=& g_{\chi_{c1} D \bar{D}^{*}} \epsilon_{\chi_{c1}}\cdot \epsilon^*_{{D}^{*0}}.
\end{eqnarray}

The quantum numbers of  $D_s^+\pi^-$ system in relative $S$- and $P$-wave are $J^P=0^+$ and $1^-$, respectively. For the rescattering processes in Fig.~\ref{feynman-diagram}(a/b), we are interested in the near-threshold $S$-wave $D^{*0} K^{*0}$/$D_s^{*+}\rho^-$ scattering into $D_s^+\pi^-$. The quantum numbers of $D^{*0} K^{*0}$/$D_s^{*+}\rho^-$ system in relative $S$-wave can be $J^P=0^+$, $1^+$ and $2^+$. By taking into account requirements of the parity and angular momentum conservation, the scattering amplitude for $D^{*0} K^{*0} \to D_s^+\pi^-$ can be written as
\begin{eqnarray}\label{DstarKstartoDspi}
	\mathcal{A}(D^{*0} K^{*0} \to D_s^+\pi^-) &=& C_{a} \epsilon_{D^{*0}}\cdot \epsilon_{K^{*0}},
\end{eqnarray}
where $C_a$ is the coupling constant.

The decay amplitude of $B^0\to \bar{D}^0 D_s^+ \pi^-$ via the $\chi_{c1}K^{*}D^{*}$ loop in Fig.~\ref{feynman-diagram} (a) is given by
\begin{eqnarray}\label{amp-loop-a}
	&&\mathcal{A}_{B^0\to \bar{D}^0 D_s^+ \pi^-}^{[ \chi_{c1}K^{*}D^{*} ]} = -{i} \int \frac{d^4q_1}{(2\pi)^4} \frac{\mathcal{A}(B^0\to \chi_{c1}K^{*0}) }{ (q_1^2-M_{{K}^*}^2 +i M_{{K}^*}\Gamma_{{K}^*})  }  \nonumber \\
	&&\times \frac{ \mathcal{A}(\chi_{c1}\to \bar{D}^0 D^{*0}) \mathcal{A}(D^{*0} K^{*0} \to D_s^+ \pi^-) }{ (q_2^2-M_{\chi_{c1}}^2 +i M_{\chi_{c1}}\Gamma_{\chi_{c1}}) (q_3^2-M_{D^{*}}^2) } ,
\end{eqnarray}
where the sum over polarizations of intermediate state is implicit. For the intermediate spin-1 state, the sum over polarization takes the form $\sum_{\mbox{pol}} \epsilon_\mu \epsilon_\nu^*=-g_{\mu\nu}+v_\mu v_\nu$, and we set $v=(1,\boldsymbol{0})$ for a non-relativistic approximation.  The Breit-Wigner type propagators are introduced in Eq.~(\ref{amp-loop-a}) to account for the width effects of intermediate states. The loop integral is performed by employing the program package \textit{LoopTools}~\cite{Hahn:1998yk}.

There are several experimentally established $\chi_{c1}$ states, of which the masses are close to TS region shown in Table~\ref{table:TSregion}. We use the latest LHCb results~\cite{LHCb:2021uow}:
\begin{eqnarray}\label{eq:mass-chic1}
	\chi_{c1}(4140): && 	M\simeq 4118\ \mbox{MeV},
	\Gamma\simeq 162~\  \mbox{MeV},  \nonumber\\
	\chi_{c1}(4274): && M\simeq 4294\ \mbox{MeV},
	\Gamma\simeq 53~\  \mbox{MeV},   \nonumber\\
	\chi_{c1}(4685): && M\simeq 4684\ \mbox{MeV},
	\Gamma\simeq 126~\  \mbox{MeV}.   
\end{eqnarray}

We call the TS peak induced by the rescattering process as the signal. There is also background for the $B^0\to \bar{D}^0 D_s^+ \pi^-$ decays. In our context, the background is defined as other contributions which are not from the rescattering diagrams of Fig.~\ref{feynman-diagram}. For $B^0\to \bar{D}^0 D_s^+ \pi^-$ decays, the experiment shows the important contributions are from the decay chain  $B^0\to D_s^+ \bar{D}_J \to  D_s^+ \bar{D}^0\pi^-$, where $\bar{D}_J$ represents an anti-charmed meson that can decay into $\bar{D}^0\pi^-$~\cite{LHCb:Tcs2900}. As a rough estimation, we parametrize the background amplitude as follows
\begin{eqnarray}\label{eq:bk}
	\mathcal{A}_{bk}=b_0 e^{i\theta_0} T_{S-wave}+b_1 e^{i\theta_1} T_{D^*}+b_2 e^{i\theta_2} T_{D_2},
\end{eqnarray}
where the coefficients $b_j e^{i\theta_j}$ ($j$=0, 1, 2) describe the relative contribution of each intermediate process. The $T_{S-wave}$, $T_{D^*}$ and $T_{D_2}$ represent the $\bar{D}^0\pi^-$ $S$-wave, $D^{*}(2010)^-$ and $D_2(2460)$ contributions respectively, which are parametrized by the relativistic Breit-Wigner amplitudes together with the angular distribution functions~\cite{LHCb:Tcs2900}. These three contributions are dominant in the $B^0\to \bar{D}^0 D_s^+ \pi^-$ decays, of which the fit fractions are around $45.0\%$, $17.0\%$ and $22.35\%$, respectively. The fit fraction of $T_{c\bar{s}}(2900)$ in LHCb experiment is about $2.55\%$. The branching fraction of $B^0\to D_s^+ D^*(2010)^-$ is given to be $(8.0\pm 1.1)\times 10^{-3}$ in Ref.~\cite{PDG:2022}. We can estimate the following ratio using the LHCb fit fraction results~\cite{LHCb:Tcs2900}
\begin{eqnarray}
	&&\frac{\mathcal{B}(B^0\to \bar{D}^0 T_{c\bar{s}}(2900))\times \mathcal{B}(T_{c\bar{s}}(2900)\to D_s^+\pi^-)}{\mathcal{B}(B^0\to D_s^+ D^{*-})\times \mathcal{B}(D^{*-}\to \bar{D}^0\pi^-)} \nonumber \\
	&&=\frac{(2.55\pm 0.93)\%}{(17.6\pm 2.6)\%} =0.15\pm 0.06,
\end{eqnarray}
which further gives that the $\mathcal{B}(B^0\to \bar{D}^0 T_{c\bar{s}}(2900))\times \mathcal{B}(T_{c\bar{s}}(2900)\to D_s^+\pi^-)$ is around $(0.81\pm 0.34)\times 10^{-3}$.
Taking into account the above background, the complete amplitude of $B^0\to \bar{D}^0 D_s^+ \pi^-$ is then given by 
\begin{eqnarray}\label{eq:totalamp}
	\mathcal{A}=\mathcal{A}_{bk} + e^{i\phi} \mathcal{A}^{loop},
\end{eqnarray}
where $e^{i\phi}$ describes the relative phase between the background and rescattering amplitude with the TS signal involved.

In order to give the numerical estimation of the invariant mass spectrum, we need to estimate the coupling constants in relevant. Concerning the $\chi_{c1}K^* D^*$ diagram, unfortunately the three vertices are not well known. The experimental data of the $B^0$ decaying into a $K^{*0}$ and a higher $\chi_{c1}$ state is not available yet. We assume the branching fraction $\mathcal{B}(B^0\to K^{*0} \chi_{c1}(4274))$ is about $10^{-3}$. Then we have $|w_a|^2/\Gamma_{B^0}\simeq 0.63$ GeV for $\chi_{c1}(4274)$ diagram. For the other $\chi_{c1}$ diagrams, we naively use the same weak coupling as that of $\chi_{c1}(4274)$. One consideration is that the $K^*\chi_{c1}(4685)$ threshold is larger than $M_{B^0}$. We set the partial decay widths of $\chi_{c1}\to \bar{D}^0 D^{*0}$ to be 20 MeV, 10 MeV and 20 MeV for $\chi_{c1}(4140)$, $\chi_{c1}(4274)$ and $\chi_{c1}(4685)$, respectively, taking into account their different total widths as shown in Eq.~(\ref{eq:mass-chic1}). The coupling $g_{\chi_{c1} D \bar{D}^{*}}$ is then estimated according to the partial width. For the contact interaction $D^{*0} K^{*0} \to D_s^+\pi^-$, we take the value $c_a\simeq 62$, and fix the relative phase $\phi\simeq -2.33$ in Eq.~(\ref{eq:totalamp}). Using these couplings the contribution of the interference term $2{Re}(\mathcal{A}_{bk}^* e^{i\phi} \mathcal{A}^{loop})$  between the background and $\chi_{c1}(4274)$ loop amplitude is comparable to that of the $T_{c\bar{s}}(2900)$ resonance. After integrating over the phase space, the branching fraction of $B^0\to \bar{D}^0 D_s^+ \pi^-$ given by the constructive interference term is around $1.0\times 10^{-3}$. For the complex coefficients in $\mathcal{A}_{bk}$, we input the experimental fitting results in the calculations~\cite{LHCb:Tcs2900}.

\begin{figure}[htbp]
	\centering
	\includegraphics[width=0.8\hsize]{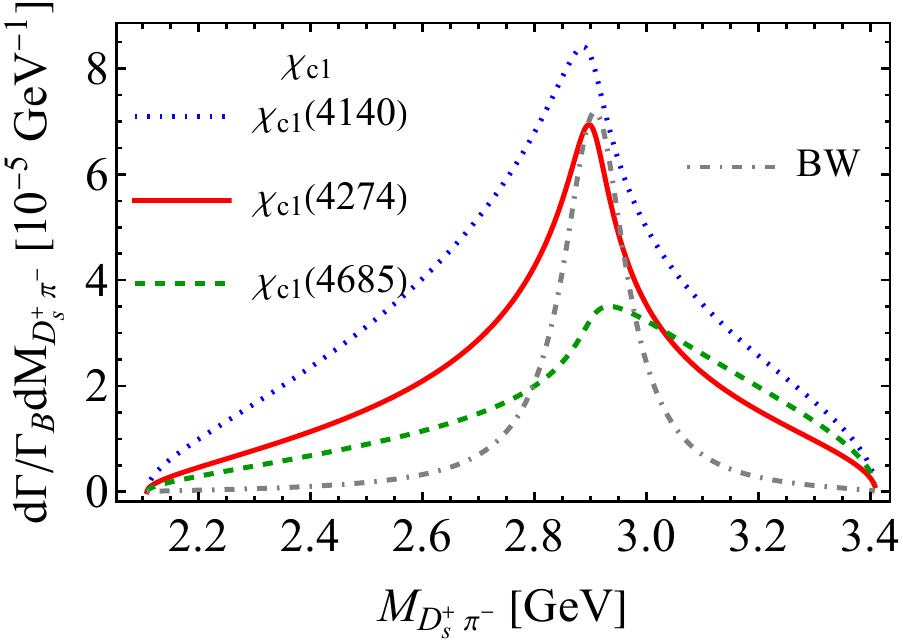}
	\caption{Invariant mass distributions of $D_s^+\pi^-$ via the rescatterings in Fig.~\ref{feynman-diagram}(a). The mass and width of intermediate state $\chi_{c1}$ are taken to be those of $\chi_{c1}(4140)$ (dotted line), $\chi_{c1}(4274)$ (solid line),  and $\chi_{c1}(4685)$ (dashed line) given in Eq.~(\ref{eq:mass-chic1}), separately. The dot-dashed line indicates the $T_{c\bar{s}}(2900)$ Breit-Wigner resonance curve for comparation. }\label{mDspi:chic1}
\end{figure}

The numerical results of the $D_s^+ \pi^-$ invariant mass distributions via the rescattering process of Fig.~\ref{feynman-diagram}(a) are displayed in Fig.~\ref{mDspi:chic1}. In order to compare with the resonance lineshape, a distribution curve corresponding to the $T_{c\bar{s}}(2900)$ is also shown in the plot, where a Breit-Wigner amplitude is employed in parameterizing the resonance amplitude. From Fig.~\ref{mDspi:chic1}, one can see that the rescattering amplitude generates a peak around 2.9 GeV. Especially for the $\chi_{c0}(4274)$ diagram, the distribution curve is comparable with the $T_{c\bar{s}}(2900)$ resonance curve to some extent. On the other hand, the curves corresponding to the $\chi_{c1}(4140)$ and $\chi_{c1}(4685)$ diagrams are broader.  This is because the TS location of  the $\chi_{c0}(4274)$ diagram is much closer to the physical boundary as shown in Fig.~\ref{TS-position}. It is possible that the TS peak may mimic the resonance-like structure in the $D_s^+\pi^-$ spectrum around the $D^*K^*$ threshold. However, even for the $\chi_{c0}(4274)$ diagram, one may notice that the TS peak is not quite narrow. This is because we also take into account the width effect of the intermediate $K^*$ state in Eq.~(\ref{amp-loop-a}).

\subsection{$D^{**} D_s^* \rho$ loop}

For the $B^0\to \bar{D}^0 D_s^+ \pi^-$ decay via the $D^{**} D_s^* \rho$ rescattering diagram as shown in Fig.~\ref{feynman-diagram}(b), the relevant amplitudes read
\begin{eqnarray}
		\mathcal{A}(B^0\to D_s^{*+}D^{**-}) &=& w_b \epsilon^*_{D_s^{*+}}\cdot \epsilon^*_{D^{**-}}, \\
		\mathcal{A}(D^{**-}\to \bar{D}^0 \rho^-) &=& g_{D^{**}D\rho} \epsilon_{D^{**-}}\cdot \epsilon^*_{\rho^-}, \label{eq:DstarstarRho}\\
		\mathcal{A}(D_s^{*+} \rho^- \to D_s^+\pi^-) &=& C_{b} \epsilon_{D_s^{*+}}\cdot \epsilon_{\rho^-},
\end{eqnarray}
where $w_b$, $g_{D^{**}D\rho}$ and $C_{b}$ represent the coupling constants. The rescattering amplitude takes the form
\begin{eqnarray}\label{amp-loop-b}
	&&\mathcal{A}_{B^0\to \bar{D}^0 D_s^+ \pi^-}^{[ D^{**} D_s^* \rho  ]} = -{i} \int \frac{d^4q_1}{(2\pi)^4} \frac{\mathcal{A}(B^0\to D_s^{*+}D^{**-}) }{ (q_1^2-M_{D_s^*}^2 )  } \times  \nonumber \\
	&& \frac{ \mathcal{A}(D^{**-}\to \bar{D}^0 \rho^-) \mathcal{A}(D_s^{*+} \rho^- \to D_s^+\pi^-) }{ (q_2^2-M_{D^{**}}^2 +i M_{D^{**}}\Gamma_{D^{**}}) (q_3^2-M_{\rho}^2 +i M_\rho \Gamma_\rho) } ,
\end{eqnarray}
where the width impacts of $D^{**}$ and $\rho$ mesons are taken into account employing the Breit-Wigner propagators.

There are some $D^{**}$ candidates of which the masses are in the vicinity of TS region shown in Table~\ref{table:TSregion}.  The LHCb collaboration has reported two states around 3000 MeV, the natural parity state $D_J^*(3000)$ state with $M=3008.1\pm 4.0$ MeV and $\Gamma=110.5\pm 11.5$ MeV and the unnatural parity state $D_J(3000)$ with $M=2971.8\pm 8.7$ MeV and $\Gamma=188.1\pm 44.8$ MeV~\cite{LHCb:2013jjb}. The $D_J(3000)$ is found in the $D^{*}\pi^-$ spectrum, of which the quantum numbers could be $1^+$. In the quark model classifications, the $D_J(3000)$ favors the $D(2P_1)$ or $D(2P_1^\prime)$ assignment~\cite{Godfrey:2015dva,Ni:2021pce,Ebert:2009ua,Li:2010vx,Zeng:1994vj}. The physical states $D(2P_1)$ and $D(2P_1^\prime)$ are usually understood as the mixed states between $2^1P_1$ and $2^3P_1$ states. In the following numerical calculations, we employ the quark model results of Ref.~\cite{Godfrey:2015dva}:
\begin{eqnarray}\label{eq:mass-Dstarstar}
	D(2P_1): && 	M= 2924\ \mbox{MeV},
	\Gamma\simeq 125~\  \mbox{MeV},  \nonumber\\
	&& \Gamma_{D\rho}=3.4\  \mbox{MeV},  \nonumber\\
	D(2P_1^\prime): && M= 2961\ \mbox{MeV},
	\Gamma= 212~\  \mbox{MeV},    \nonumber\\
	&& \Gamma_{D\rho}=18.8\  \mbox{MeV},
\end{eqnarray}
where the partial decay width of the $D\rho$ channel is also given, which can be used to determine the coupling constant $g_{D^{**}D\rho}$ in Eq.~(\ref{eq:DstarstarRho}). For the experimental observed state $D_J(3000)$, we assume the partial decay width of the $D\rho$ channel is about 20 MeV.

For the Cabibbo-favored weak decay $B^0\to D_s^{*+}D^{**-}$, we assume the branching fraction is at the order of $10^{-3}$. Then we have $|w_b|^2/\Gamma_{B^0}\simeq 0.57$ GeV for the $D_J(3000)$ diagram.
For the contact interaction, we take the value $c_b\simeq 43$, and fix the relative phase $\phi\simeq -2.16$. Using these couplings, after integrating over the phase space, the branching fraction of $B^0\to \bar{D}^0 D_s^+ \pi^-$ given by the constructive interference term is around $1.0\times 10^{-3}$, which is comparable to the contribution of the $T_{c\bar{s}}(2900)$ resonance.

\begin{figure}[htbp]
	\centering
	\includegraphics[width=0.8\hsize]{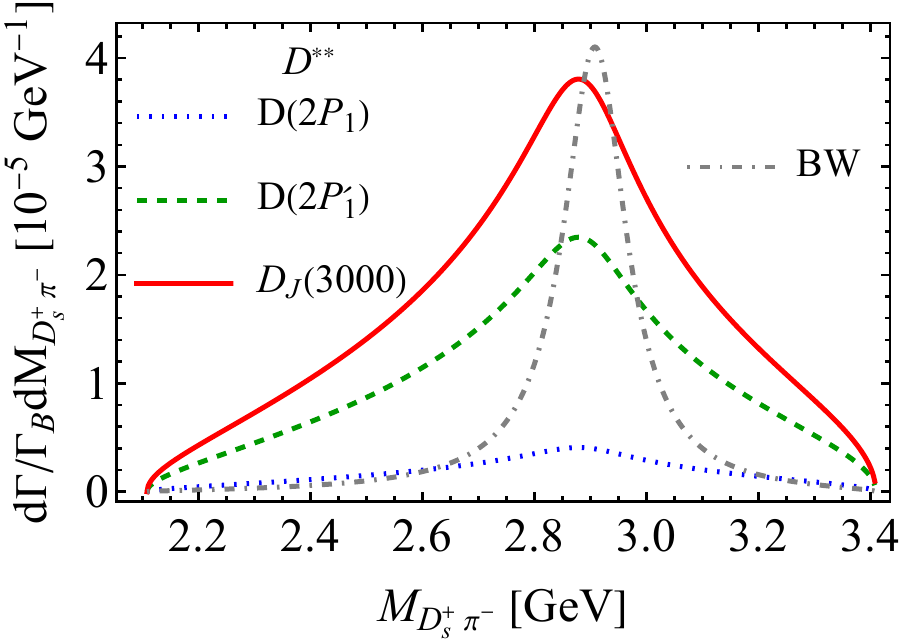}
	\caption{Invariant mass distributions of $D_s^+\pi^-$ via the rescatterings in Fig.~\ref{feynman-diagram}(b). The mass and width of intermediate state $D^{**}$ are taken to be those of $D(2P_1)$ (dotted line), $D(2P_1^\prime)$ (dashed line),  and $D_J(3000)$ (solid line), separately. The dot-dashed line indicates the rescaled $T_{c\bar{s}}(2900)$ Breit-Wigner resonance curve for comparation. }\label{mDspi:D1}
\end{figure}

The invariant mass distribution curves of $D_s^+\pi^-$ via the $D^{**} D_s^* \rho$ diagrams are shown in Fig.~\ref{mDspi:D1}.
Although the mass of $D(2P_1)$, $D(2P_1^\prime)$ or $D_J(3000)$ perfectly satisfies the TS kinematic conditions as shown in Table~\ref{table:TSregion}, the pertinent invariant mass distribution curves are still much broader compared with the resonance lineshape. This is mainly because the intermediate state $\rho$ is very broad. The PDG gives $\Gamma_\rho=149$ MeV~\cite{PDG:2022}, and we have taken into account this broad width in the loop integral as shown in Eq.~(\ref{amp-loop-b}). Therefore the TS peak itself corresponding to the $D^{**} D_s^* \rho$ diagram can hardly simulate the relatively narrower resonance-like structure $T_{c\bar{s}}(2900)$, although such rescattrings can still contribute to the $B^0\to \bar{D}^0 D_s^+ \pi^-$ decays.

\subsection{$\chi_{c0}K D $ loop}

There have been many theoretical studies concerning the Nambu-Goldstone-bosons scattering off the heavy flavor mesons. In the $(S,\ I)=(1,\ 0)$ channel, the scattering length $a_{DK}^{I=0}$ is generally predicted to be large. By means of lattice QCD simulations and chiral extrapolation, in Ref.~\cite{Liu:2012zya} the authors give $a_{DK}^{I=0}=(-0.86\pm 0.03)\ \mbox{fm}$ at the physical pion mass. The large negative scattering length $a_{DK}^{I=0}$ indicates the isoscalar $DK$ interaction is strong. Furthermore, it is generally supposed that the $D_{s0}^{*}(2317)$/$D_{s1}(2460)$ is the hadronic molecule dynamically generated by the strong $DK$/$D^*K$ ($I=0$) interaction in the coupled-channels dynamics \cite{Guo:2006fu,Guo:2009ct,Liu:2009uz,Kolomeitsev:2003ac,Altenbuchinger:2013vwa,Guo:2015dha,Liu:2012zya,Mohler:2013rwa,Chen:2016spr,Yao:2015qia}. On the other hand, the isovector $DK$ interaction is usually though to be relatively weak, and the existence of isovector dynamically generated resonant or bound states composed of $D^{(*)}K$/$\bar{B}^{(*)}K$ is generally not expected. If one observes some resonance-like structures around the $D^{(*)}K$/$\bar{B}^{(*)}K$ threshold in the $(S,\ I)=(1,\ 1)$ channel, such as the $D^{(*)}_s\pi$ or $B^{(*)}_s\pi$ channel, it is very likely these structures may have some other origins, such as the threshold effects. Searching for such exotic resonance-like structures in the $D_s\pi$ and $B_s\pi$ spectrum has ever been proposed in Refs.~\cite{Liu:2015taa,Liu:2017vsf}.

In the recent LHCb measurement on the $B\to \bar{D} D_s\pi$ decays, there is no obvious peak structure around the $DK$ threshold observed in the $D_s\pi$ spectrum.
Similar to the above two subsections, we hope to find out whether the rescattering process in Fig.~\ref{feynman-diagram}(c) may play a role, especially whether the TS peak around the $DK$ threshold generated from the rescattering diagram can be observed.

For the $B^0\to \bar{D}^0 D_s^+ \pi^-$ decay via the $\chi_{c0}K D$ rescattering diagram as shown in Fig.~\ref{feynman-diagram}(c), we introduce the following $S$-wave couplings
\begin{eqnarray}
	\mathcal{A}(B^0\to \chi_{c0}K^{0}) &=& w_c ,\\
	\mathcal{A}(\chi_{c0}\to \bar{D}^0 D^0)&=&g_{\chi_{c0} D\bar{D}}.
\end{eqnarray}
The rescattering amplitude takes the form
\begin{eqnarray}\label{amp-loop-c}
	&&\mathcal{A}_{B^0\to \bar{D}^0 D_s^+ \pi^-}^{[ \chi_{c1}KD ]} = -{i} \int \frac{d^4q_1}{(2\pi)^4} \frac{\mathcal{A}(B^0\to \chi_{c0}K^{0}) }{ (q_1^2-M_{{K}}^2)  }  \nonumber \\
	&&\times \frac{ \mathcal{A}(\chi_{c0}\to \bar{D}^0 D^{0}) \mathcal{A}(D^{0} K^{0} \to D_s^+ \pi^-) }{ (q_2^2-M_{\chi_{c0}}^2 +i M_{\chi_{c0}}\Gamma_{\chi_{c0}}) (q_3^2-M_{D}^2) },
\end{eqnarray}
where the amplitude $\mathcal{A}(D^{0} K^{0} \to D_s^+ \pi^-) $ is discussed in detail below.

There are several established higher $\chi_{c0}$ states above the $D\bar{D}$ threshold, i.e., $\chi_{c0}(3915)$, $\chi_{c0}(4500)$ and $\chi_{c0}(4700)$.
We use the parameters from PDG 2022~\cite{PDG:2022} and the LHCb experiment~\cite{LHCb:2021uow}:
\begin{eqnarray}\label{eq:mass-chic0}
	\chi_{c0}(3915): && 	M\simeq 3922\ \mbox{MeV},
	\Gamma\simeq 19~\  \mbox{MeV},  \nonumber\\
	\chi_{c0}(4500): && M\simeq 4474\mbox{MeV},
	\Gamma\simeq 77~\  \mbox{MeV},   \nonumber\\
	\chi_{c0}(4700): && M\simeq 4694\ \mbox{MeV},
	\Gamma\simeq 87~\  \mbox{MeV}.  
\end{eqnarray}
The $\chi_{c0}(3915)$ state is observed in the $D\bar{D}$ spectrum, which is generally though to be the excited charmonium state $\chi_{c0}(2P)$. The $\chi_{c0}(4500)$ (aka $X(4500)$ )  and $\chi_{c0}(4700)$ (aka $X(4700)$ ) have been found in the $J/\psi\phi$ spectrum~\cite{LHCb:2021uow,LHCb:2016axx}, and their nature is still unclear. Anyway, since they are chamronium-like states with $J^{PC}=0^{++}$, it is expect that they can decay into $D\bar{D}$ states in relative $S$-wave. We set the moderate partial decay widths of $\chi_{c0}\to \bar{D}^0 D^{0}$ to be 10 MeV, 20 MeV and 20 MeV for $\chi_{c0}(3915)$, $\chi_{c0}(4500)$ and $\chi_{c0}(4700)$, respectively. The coupling $g_{\chi_{c0} D\bar{D}}$ then can be estimated. For the weak decay $B^0\to \chi_{c0}K^{0}$, we assume the branching fraction is at the order of $10^{-3}$. The coupling for $\chi_{c0}(4700)$ is estimated to be $|w_c|^2/\Gamma_{B^0}\simeq 2.4$ GeV.

For the vertex $D^{0} K^{0} \to D_s^+ \pi^-$ in Fig.~\ref{feynman-diagram}(c), we employ the amplitude which is unitarized according to the method of UChPT ~\cite{Oller:2000fj,Oller:2000ma,Oller:1997ng}.
We consider the $S$-wave $DK$-$ D_s\pi$ coupled-channel scattering. The unitary $T$-matrix is given by $T=(1-VG)^{-1}V$, where $V$ represents the $S$-wave driving potential, and $G$ is a diagonal matrix composed of two-meson-scalar-loop functions \cite{Oller:2000fj,Oller:2000ma,Oller:1997ng}. In the numerical calculation, the next-to-leading-order potential from Ref.~\cite{Altenbuchinger:2013vwa} is employed, where the pertinent low-energy-constants and subtraction constant are determined by fitting the lattice QCD result of Ref.~\cite{Liu:2012zya}. We also suggest Refs.~\cite{Guo:2009ct,Guo:2015dha,Liu:2012zya,Altenbuchinger:2013vwa,Oller:2000fj,Yan:2018zdt} for more details about the formulation of the Nambu-Goldstone-bosons scattering off the heavy hadrons.

\begin{figure}[htbp]
	\centering
	\includegraphics[width=0.8\hsize]{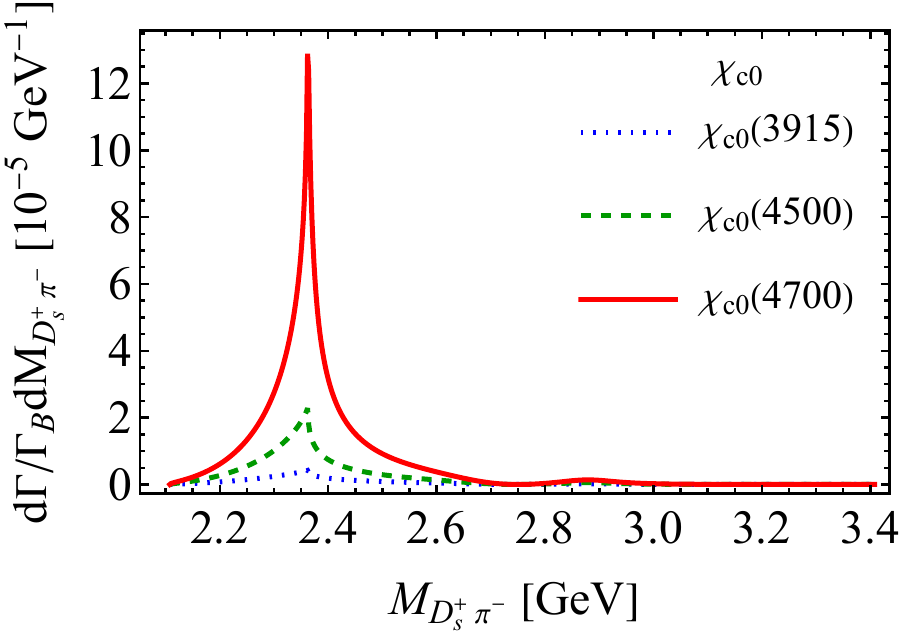}
	\caption{Invariant mass distributions of $D_s^+\pi^-$ via the rescatterings in Fig.~\ref{feynman-diagram}(c) without taking into account the interference with the background. The masses and widths of different $\chi_{c0}$ listed in Eq.~(\ref{eq:mass-chic0}) are adopted.}\label{mDspi:chic0}
\end{figure}

\begin{figure}[htbp]
	\centering
	\includegraphics[width=0.8\hsize]{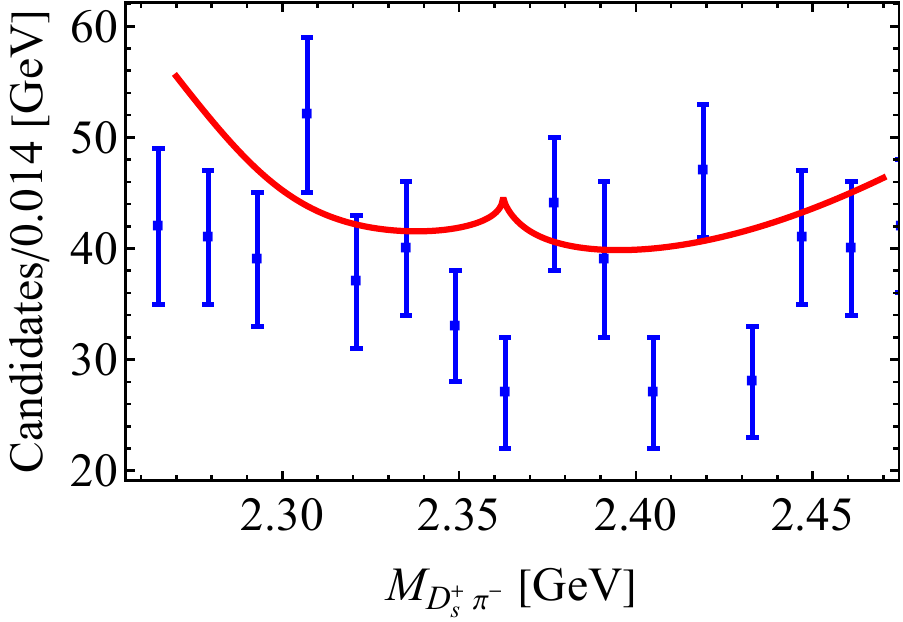}
	\caption{Invariant mass distributions of $D_s^+\pi^-$ via the rescatterings in Fig.~\ref{feynman-diagram}(c) with the interference contribution involved. The data with errors are from Ref.~\cite{LHCb:Tcs2900}. }\label{mDspi:data}
\end{figure}

The $D_s^+ \pi^-$ invariant mass distribution of the $B^0\to \bar{D}^0 D_s^+ \pi^-$ decay via the rescattering process of Fig.~\ref{feynman-diagram}(c) is displayed in Fig.~\ref{mDspi:chic0}. We can see resonance-like peaks appear in the vicinity of $DK$ threshold ($\sim 2362$ MeV). Especially for the $\chi_{c0}(4700)$ diagram, the corresponding peak is quite sharp. This is because the mass of $\chi_{c0}(4700)$ perfectly falls into the TS kinematic region as shown in Table~\ref{table:TSregion}. Another difference between the rescatterings of $\chi_{c0}K D $ loop and $\chi_{c1}K^* D^*$/$D^{**} D_s^* \rho$ loop is that both the $D^0$ and $K^0$ intermediate states are relatively stable. The TS peak is not smoothed by the width effects. 

The distributions curves shown in Fig.~\ref{mDspi:chic0} are the results by only taking into account the rescattering process itself. Although the TS peak is very obvious, its contribution may be submerged in the larger background. In terms of Eq.~(\ref{eq:totalamp}), we give a simulation of the distribution curve around 2.36 GeV, and compare it with the data, as shown in Fig.~\ref{mDspi:data}. In the simulation, we only take into account the $\chi_{c0}(4700)$ diagram and fix the relative phase $\phi=1.2$ in Eq.~(\ref{eq:totalamp}), which gives a constructive interference. From Fig.~\ref{mDspi:data}, one can see that a small peak grows up on the background. To observe such phenomena, more accurate experimental data are necessary.

\section{Summary}

In summary, we investigate the $B\to \bar{D}D_s \pi$ decay via the $\chi_{c1}K^* D^*$, $D^{**} D_s^* \rho$ and $\chi_{c0}K D $ intermediate rescattering processes. It is shown that the kinematic conditions for the TS of rescttering amplitudes locating close to the physical boundary can be well satisfied in some cases. The TS peak around the $D^*K^*$ threshold generated from the $\chi_{c1}K^* D^*$ loop is relatively narrow, which can simulate the resonance-like structure $T_{c\bar{s}}(2900)$ observed in the $D_s\pi$ spectrum. However, the TS peak around the $D_s^*\rho$ threshold generated from the $D^{**} D_s^* \rho$ is smoothed by the broad width of $\rho$, which itself can hardly describe the $T_{c\bar{s}}(2900)$ structure, but it still can play a role in the $B\to \bar{D}D_s \pi$ decay. In conclusion, it is possible that the fully open-flavor tetraquark candidates $T_{c\bar{s}}(2900)$ and $X(2900)$ can be interpreted in the same picture, i.e., both of them may be resulted from the threshold effects. However, we should also mention that some couplings in the rescatterings discussed here are not well know yet. Although the presence of TS peaks mainly depend on the kinematics, the relative strength between the TS peak and background still quite depend on these couplings.  

The resonance-like structure around $DK$ threshold in the $D_s\pi$ spectrum is also studied. It is shown that a small but narrow TS signal generated from the $\chi_{c0}K D $ rescattering diagram may grow on the background. Due to relative weak $DK$ ($I=1$) interaction, no dynamic pole is expected around 2.36 GeV. If one observes such a structure in the $D_s\pi$ distribution, we can conclude that it may results from the TS. More accurate measurement is highly recommended in future experiments.

\begin{acknowledgments}
We thank Q. Zhao for helpful discussions. This work is supported, in part, by the National Natural Science Foundation of China (NSFC) under Grant Nos.~11975165 and 12075167. 
\end{acknowledgments}

\end{document}